# Directional Interlayer Spin-Valley Transfer in Two-Dimensional Heterostructures

**Authors**: John R. Schaibley[1], Pasqual Rivera[1], Hongyi Yu[2], Kyle L. Seyler[1], Jiaqiang Yan[3-4], David G. Mandrus[3-5], Takashi Taniguchi[6], Kenji Watanabe[6], Wang Yao[2], and Xiaodong Xu[1,7]

**Affiliations:**

[1]Department of Physics, University of Washington, Seattle, Washington 98195, USA

[2]Department of Physics and Center of Theoretical and Computational Physics, University of Hong Kong, Hong Kong, China

[3]Materials Science and Technology Division, Oak Ridge National Laboratory, Oak Ridge, Tennessee, 37831, USA

[4]Department of Materials Science and Engineering, University of Tennessee, Knoxville, Tennessee, 37996, USA

[5]Department of Physics and Astronomy, University of Tennessee, Knoxville, Tennessee, 37996, USA

[6]Advanced Materials Laboratory, National Institute for Materials Science, Tsukuba, Ibaraki 305-0044, Japan

[7]Department of Materials Science and Engineering, University of Washington, Seattle, Washington 98195, USA

Correspondence to: xuxd@uw.edu

**Abstract:** Van der Waals heterostructures formed by two different monolayer semiconductors have emerged as a promising platform for new optoelectronic and spin/valleytronic applications. In addition to its atomically thin nature, a two-dimensional semiconductor heterostructure is distinct from its three-dimensional counterparts due to the unique coupled spin-valley physics of its constituent monolayers. Here, we report the direct observation that an optically generated spin-valley polarization in one monolayer can be transferred between layers of a two-dimensional $MoSe_2$-$WSe_2$ heterostructure. Using nondegenerate optical circular dichroism spectroscopy, we show that charge transfer between two monolayers conserves spin-valley polarization and is only weakly dependent on the twist angle between layers. Our work points to a new spin-valley pumping scheme in nanoscale devices, provides a fundamental understanding of spin-valley transfer across the two-dimensional interface, and shows the potential use of two-dimensional semiconductors as a spin-valley generator in 2D spin/valleytronic devices for storing and processing information.

**Introduction**

Spin initialization is a crucial operation for spintronic devices which require a net spin polarization for reading, writing, and transferring information[1]. Two-dimensional semiconductors, such as monolayer $MoSe_2$ and $WSe_2$, have recently emerged as a new spin/valleytronic platform[2-5]. Their inversion-asymmetric honeycomb lattice structures give rise to two energy degenerate but inequivalent (+K and –K) momentum-space valleys, forming a pseudospin system analogous to real electron spin[6]. Due to strong spin-orbit coupling, the valley pseudospin is locked to the real spin orientation[6]. Since flipping an electron spin requires a simultaneous flip of a valley pseudospin, free carrier spin-valley polarization at the band edge is expected to be robust and long lived, which has recently been measured to be on order of 1-100 ns[7,8]. Large spin-valley polarizations associated with excitons have been generated by circularly polarized light excitation through a valley dependent optical selection rule[3,6,9-11]. However, excitonic spin-valley polarization in a monolayer does not last long compared to free carriers due to the picosecond timescale of the valley exciton depolarization time, which arises from the electron-hole exchange interaction[3,12-14] and the ultrafast decay time of the exciton itself[15,16]. In addition, it is not clear how to exploit a monolayer system as a spin generator to supply optically generated spin-valley polarization to a different physical system.

2D semiconductor heterostructures formed by stacking two monolayers on top of each other can be designed to realize new spin-valley systems with important advantages over individual monolayers. It has been established that $WX_2$-$MoX_2$ (where X = S, Se) heterostructures have a type-II band alignment[17,18], which leads to ultrafast charge transfer between layers and tunable photodetectors[19,20]. Such spatial separation of electrons and holes suppresses ultrafast electron-hole recombination[21-24] and their exchange interaction[21-26], both of which limit the practical application of optical spin-valley orientation in monolayers[3,27,28]. Very recently, helicity-dependent photoluminescence (PL) measurements of interlayer excitons revealed spin-valley polarization lifetimes exceeding tens of nanoseconds[26], showing that the spatial separation of electrons and holes indeed provides a powerful approach towards practical spin-valleytronics. However, interlayer exciton effects were accompanied by complicated electron-hole relaxation pathways and the effect of the twist angle between the two layers[25], which complicate the quantitative analysis of spin-valley polarization from the polarization resolved interlayer exciton PL. In addition, interlayer exciton PL studies were limited to small twist angle samples only, because the electron-hole momentum mismatch in large twist angle heterostructures strongly suppresses interlayer exciton light emission. All of these limitations obscured a clear understanding of the unique spin-valley properties of 2D semiconductor heterostructures, especially the transport of spin-valley polarized free carriers across the 2D layer interface.

In this work, by applying polarization resolved non-degenerate nonlinear optical spectroscopy, we provide a direct probe of interlayer spin-valley polarization transfer in a model 2D heterostructure with varying twist angles formed by monolayer $MoSe_2$ and $WSe_2$. By optically exciting an intralayer exciton spin-valley polarization in one layer and probing the intralayer neutral and charge excitons in different layers, we demonstrate that the subsequent interlayer charge transfer is directional and conserves spin, i.e. spin polarization transfer leads to polarized hole spins in $WSe_2$ and electron spins in $MoSe_2$ (Fig. 1a). We find that the spin-valley polarization transfer has only a weak dependence on twist angles in the heterobilayer. Our results realize directional pumping of spin-valley polarized carrier spins into individual layers of a 2D heterostructure by harnessing the coupled spin-valley physics of the constituent monolayers[6].

**Results**

**Sample fabrication and electronic structure.** The $MoSe_2$-$WSe_2$ heterostructures were fabricated from independently isolated, exfoliated monolayers (see Fig. 1b). In order to investigate the effect of heterostructure twist angle, we first measured the crystal axes of individual monolayers by polarization-resolved and phase-sensitive second-harmonic generation spectroscopy[29-32] (see Supplementary Figure 1 and Supplementary Note 1). The monolayers were then assembled into heterostructures using a dry transfer stamping technique[33] with known twist angle. Results from heterostructures with non-zero twist angels are presented in Supplementary Note 2 and Supplementary Figures 2-3. The sample in the main text has a twist angle near 0º, where the valleys from the different layers are nearly aligned in momentum space (Fig. 1c). The lowest conduction band is located in the $MoSe_2$ and the highest valence band in $WSe_2$. Within each monolayer, $\sigma \pm$ circularly polarized light couples to transitions in the $\pm$ K valley only. The high quality of our heterostructure was confirmed by observing a strong PL quenching of the intralayer excitons, and the observation of interlayer excitons (see Supplementary Figure 4), where Coulomb-bound electrons and holes are localized in opposite layers[24].

**Nonlinear excitonic response of the heterostructure.** We first determined the energy position of intralayer excitons by performing energy resolved continuous wave differential transmission (DT) or differential reflection (DR) spectroscopy[34]. This is a two beam pump-probe technique which measures the difference of the probe transmission or reflection when the pump is on and off. The experiments in the main text were all performed on the same heterostructure mounted on sapphire. Additional measurements were performed on different heterostructures on $SiO_2$ substrates and are in qualitative agreement with the data presented in the main text (see Supplementary Note 2). The experiments were performed at 30 K, unless otherwise specified.

The degenerate DT spectrum of a heterostructure is shown in Fig. 1d with cross-circularly polarized pump and probe. Compared to the DT spectrum from individual monolayers, we see the intralayer exciton resonances in heterostructure are consistent with the spectral positions of isolated monolayers with a ~20 meV redshift and broader linewidth. We attribute the ~20 meV red shift to a reduction in the intralayer exciton bandgaps due to the coupling between layers. The linewidth broadening is attributed to the charge transfer between the layers, which leads to an extra relaxation channel for the intralayer excitons[35]. The resonance line shapes of the degenerate DT spectrum consist of a pump-induced increase to the probe transmission at high energy and a pump induced absorption at low energy. Note that the low energy pump induced absorption feature is stronger for the $MoSe_2$ layer compared to $WSe_2$. We attribute this difference to the different oscillator strengths of different charged exciton species in each layer (see Supplementary Note 3).

**Demonstration of interlayer charge transfer.** To establish interlayer carrier transfer, we performed two-color nondegenerate DR and DT measurements. Both types of measurements were performed on the same sample and the data are qualitatively similar. We use the DT data exclusively in curve fitting to avoid the interference effects that arise from the substrate reflection in the DR measurements. Fig. 2a shows the DR spectrum with co-circularly polarized pump and probe, where the pump is resonant with the lower energy $MoSe_2$ exciton at 1.621 eV while the probe laser scans over the $WSe_2$ exciton resonance near 1.68 eV. The green curve shows an enhanced DR response from the heterostructure region. In comparison, the black curve shows the DR response when both pump and probe are focused on an isolated monolayer $WSe_2$ region which shows a negligible DR response when the pump energy is fixed at the $MoSe_2$ exciton resonance. In the heterostructure, since the $MoSe_2$ exciton has lower energy than $WSe_2$, the observed DR response near the $WSe_2$ exciton when pumping the $MoSe_2$ exciton resonance is unlikely from the energy transfer from $MoSe_2$ exciton. Rather, it is a result of charge transfer from $MoSe_2$ to $WSe_2$. Specifically, the hole is transferred from the $MoSe_2$ valence band to the $WSe_2$ valence band due to the type-II band alignment.

**Demonstration of interlayer spin-valley polarization transfer.** Interlayer spin-valley transfer was then investigated by performing polarization resolved DT experiments which measure the pump induced circular dichroism (CD). The pump laser polarization and energy were chosen to only excite valley polarized excitons in the $MoSe_2$ layer. The DT spectrum was measured for both co- (burgundy curve) and cross- (green curve) circularly polarized configurations for the probe scanning through the $WSe_2$ excitons (Fig. 2b). The CD can be defined as the difference between the cross- and co-polarized DT spectra for either fixed pump or fixed probe polarization. Both yield similar results. For the convenience of our experimental configuration, we choose to fix the probe helicity while switching the pump helicity, i.e. $CD = \frac{DT}{T}(\sigma_{pump}^-) - \frac{DT}{T}(\sigma_{pump}^+)$, where the subscript denotes the pump beam, and $T$ is the probe transmission. As shown in Fig. 2c, the sign of the pump-induced CD response reverses for opposite probe helicities. The observed CD demonstrates a valley population imbalance, i.e., the creation of spin-valley polarization in $WSe_2$. We attribute this population imbalance to the pumping of polarized hole spins as depicted in Fig. 1a. Circularly polarized excitation resonantly pumps spin-valley polarized excitons in the $MoSe_2$ layer, about 60 meV below the $WSe_2$ exciton energy. The spin polarized hole then transfers to the $WSe_2$ +K valence band, which gives rise to hole spin-valley polarization in $WSe_2$ and electron spin-valley polarization in $MoSe_2$. The observation of the CD response supports this picture.

We also demonstrate electron spin transfer from the $WSe_2$ to the $MoSe_2$ layer by resonantly pumping a $WSe_2$ spin-valley polarization and probing the $MoSe_2$ excitons (Fig. 3). Similar phenomena, including the pump-induced CD is observed, whose sign depends on probe helicity. Since the $WSe_2$ exciton has higher energy than $MoSe_2$, the observed CD will have two contributions. One is due to the electron spin transfer from the $WSe_2$ to the $MoSe_2$ conduction band. The other is from the above resonance optical excitation of valley-polarized excitons directly in the $MoSe_2$. To distinguish these two effects, we measured the DR response on the heterostructure when pumping at the $WSe_2$ resonance, and then repeated the measurement on the isolated monolayer $MoSe_2$ region of the same sample (Supplementary Figure 5). We observe a 3-fold enhancement of the DR response on the heterostructure region, which shows that electron transfer from the WSe$_2$ to the MoSe$_2$ dominates the DR response. We note that since the electron and hole spin are separated in opposite layers, the exchange interaction between the electron and hole spins is strongly suppressed. This will give rise to a long polarization lifetime[26] and contributes to the enhanced DR response.

**Origin of the DT line shapes.** We now turn to the discussion of the line shapes in the non-degenerate DT measurements (Fig. 2b and Fig. 3a), which further support the picture of directional spin transfer. For simplicity, we focus on the explanation of data in Figs. 3a. Fig. 3c-f illustrate the origins of the line shapes by pumping at the WSe$_2$ exciton resonance while probing the MoSe$_2$ excitons. The DT spectra can be understood by taking the difference between the probe transmission spectrum with the pump on and off (solid orange and dashed blue curve of Figs. 3c,e). The co-polarized pump and probe (burgundy data) laser configuration is shown in the left inset of Fig. 3a. The inset depicts the pump (solid blue line) injecting +K polarized carriers in the WSe$_2$ layer and the consequent electron transfer to the +K conduction band valley in the MoSe$_2$ monolayer. The probe (dashed red line) measures the changes in transmission spectrum of the +K MoSe$_2$ excitons. Figures 3c-d depict the effects that dominate the co-polarized DT response. Because the conduction band is partially filled, phase-space filling leads to a blue shift of the transmission resonance, and the neutral exciton (X$^o$) oscillator strength is reduced (Fig. 3c). The inset to Fig. 3c depicts the DT signal calculated by taking the difference between the orange and dashed blue curves.

The cross-polarized pump and probe configuration (green data) is depicted in the right inset of Fig. 3a. Here, the pump (solid blue line) injects carriers into the –K valley of the WSe$_2$ layer, and the subsequent electron transfer to the –K valley of the MoSe$_2$. With this –K valley electron population, when the probe beam (dashed red line) excites electron-hole pairs in +K valley, negatively charged excitons (X$^-$) can form (Fig. 3e-f). The cross-polarized DT spectrum can be understood by examining Fig. 3e, which shows the pump induced changes to the cross-polarized probe transmission spectrum. Relative to the pump off case, a population of electrons in the –K valley decreases the +K cross-polarized probe transmission at X$^-$ resonance due to the increases of X$^-$ oscillator strength, and increases the transmission at the X$^o$ resonance due to the decrease of X$^o$ oscillator strength. The inset to Fig. 3e shows the corresponding cross-polarized DT spectrum. We note that the 30 meV energy separation between the peak and dip in both the cross-polarized DT spectrum (green curve of Figs. 3a) and the CD spectra (Fig. 3b) is consistent with the binding energy of X$^-$, and therefore further supports the picture of directional electron spin-valley transfer from WSe$_2$ to MoSe$_2$.

**Discussion**

We estimate the resulting spin-valley polarization of electrons in the MoSe$_2$ layer by pumping the WSe$_2$ resonance and comparing the relative co- and cross-circularly polarized DT responses of the X$^-$ in the MoSe$_2$ layer. It has been demonstrated both experimentally[4] and theoretically[36] that the X$^-$ in MoSe$_2$ is dominantly an intervalley charged exciton with the extra electron located in the lower conduction band of the opposite valley (see Supplementary Note 4 and Supplementary Figure 6). The X$^-$ formation and the magnitude of its corresponding DT signal measures the population of polarized electrons in the valley opposite the one being probed. Therefore, spin-valley polarization ($\rho$) in the MoSe$_2$ layer resulting from interlayer spin-valley transfer can be estimated by $\rho = \frac{DT/T(cross) - DT/T(co)}{DT/T(cross) + DT/T(co)}$. Since the negative MoSe$_2$ charged exciton resonance is spectrally isolated, it is not significantly influenced by signals arising from the other resonances (Fig. 3a). We fit single Lorentzians to the MoSe$_2$ DT response near X$^-$ (1.589 eV) for both co- and cross-circularly polarized pump and probe, and find that the ratio of the dip areas for the co-polarized response is 37% of the cross-polarized response (See Supplementary Figure 7). This gives a 46% electron spin-valley polarization in the MoSe$_2$ layer. This estimation is consistent with previous measurements of the interlayer exciton in helicity-dependent PL measurements, which was reported to be limited by depolarization of the monolayer exciton prior to interlayer transfer[26].

When pumping the MoSe$_2$ and probing the WSe$_2$ excitons, the charged exciton feature is also clear in the CD response (Fig. 2c). Fitting the CD spectra with a difference of two Lorentzians, we find that the energy separation between the peak and dip is approximately 19 meV, consistent with the binding energy of positively charged excitons (X$^+$) in WSe$_2$ [2,3]. This observation supports the picture of directional polarized hole spin transfer from MoSe$_2$ to WSe$_2$. However, due to the overlap of spectral features near the WSe$_2$ positively charged exciton peak, we cannot accurately compare the co- and cross-circular DT responses of X$^+$ to estimate a hole spin-valley polarization in the WSe$_2$ layer.

We also performed measurements on additional samples with varying twist angles (Supplementary Figure 2). There are fine spectral features distinct from near zero twist angle samples, which require a future systematic study. However, both the sign and signal amplitude of the CD spectra are consistent for all twist angles, which implies that spin-valley conserved interlayer charge transport is robust for different twist angles.

Our results demonstrate that spin-valley polarized carriers can be efficiently transferred between layers, providing a novel method for optically injecting long-lived and spin-valley polarized carriers in either layer of heterostructures with arbitrary twist angles. We expect this scheme could be especially useful in recent proposals that seek to use atomically thin bilayer systems for spintronic or valleytronic applications[25], or as a platform to investigate bosonic quasiparticle effects with spin structures[37].

**Methods**

**Sample Fabrication.** The heterostructures were assembled using a polycarbonate film dry transfer technique. Supplementary Note 1 contains the methods used to determine the crystal axes. The sample in the main text was encapsulated in 5-10 nm thick hexagonal boron nitride and mounted on a c-axis sapphire substrate to allow for optical transmission measurements.

**Nonlinear Optical Measurements.** The data shown in the main text were measured in a cold-finger cryostat. Two continuous-wave tunable Ti:sapphire lasers (M$^2$ SolsTiS) provided the pump and probe beams, which were each amplitude modulated with acousto-optic modulators at frequencies near 700 kHz. Both beams were actively intensity stabilized. 20 µW probe, 40 µW pump average power were used for all spectra in the main text. Polarizers and broadband waveplates were used to set the polarization of pump and probe, which were focused onto the sample with a microscope objective to a beam spot of ~ 1 µm. The transmitted light was collected by a 15 mm spherical lens that was mounted in the cold finger of the cryostat. In the DR measurements, the reflected probe was collected with the objective. The pump beam was rejected with a cross-polarized setup or with a short or long pass filter. The probe was detected with an amplified silicon photodiode. The DT or DR signal was then measured with a phase-sensitive lock-in amplifier which was locked to the difference between the pump and probe modulation frequencies. The transmitted (T) or reflected (R) probe power was measured simultaneously with the DT or DR signal while the pump was modulated and used to normalize the DT/T or DR/R response.

**Acknowledgments:**

This work is mainly supported by the Department of Energy, Basic Energy Sciences, Materials Sciences and Engineering Division (DE-SC0008145 and SC0012509). HY and WY are supported by the Croucher Foundation (Croucher Innovation Award), the RGC of Hong Kong (HKU17305914P), and the HKU ORA. JY and DM are supported by US DoE, BES, Materials Sciences and Engineering Division. K.W. and T.T. acknowledge support from the Elemental Strategy Initiative conducted by the MEXT, Japan and a Grant-in-Aid for Scientific Research on Innovative Areas "Science of Atomic Layers" from JSPS. PR and XX acknowledge support from the State of Washington funded Clean Energy Institute. XX also acknowledges a Cottrell Scholar Award, and support from the Boeing Distinguished Professorship in Physics.

**Author Contributions**:

XX and WY conceived and supervised the project. JS, PR, and KS fabricated the devices. JS performed measurements, assisted by PR. KS performed and analyzed the polarization and phase-sensitive second-harmonic generation measurements. JS, HY, XX and WY analyzed the data. JY and DGM provided and characterized the bulk $MoSe_2$ and $WSe_2$ crystals. TT and KW provided BN crystals. JS, XX, WY and HY wrote the paper. All authors discussed the results.

**Data Availability:**

The authors declare that all of the data supporting the findings of this study are available within the article and its supplementary information file.

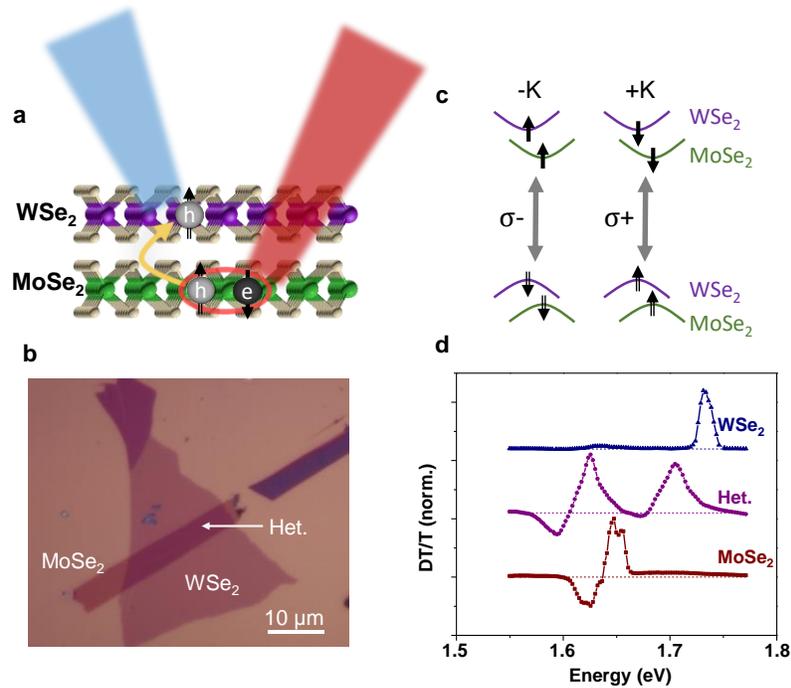

**Figure 1| Interlayer spin-valley physics. a,** Depiction of the experiment. Spin-valley polarized excitons are resonantly injected in the $MoSe_2$ layer with a polarized laser (red). The hole transfers to the $WSe_2$ layer where its spin-valley polarization is measured with another polarized laser (blue), resonant with the $WSe_2$ excitons. The black arrows depict the real spin of the electrons and holes. **b,** Optical microscope image of a $MoSe_2$-$WSe_2$ heterostructure (Het.) on $SiO_2$, showing the different sample regions. **c,** The 8-band model of the +K and –K valleys for a nearly aligned $MoSe_2$-$WSe_2$ heterostructure, showing the valley dependent optical selection rules ($\sigma\pm$ for $\pm$K valley) and real spins (black arrows) for electrons and holes. **d,** Degenerate differential transmission (DT) spectra from different sample regions for a $MoSe_2$-$WSe_2$ heterostructure on sapphire, which are normalized and stacked for comparison. The dashed lines correspond to DT/T = 0 for each spectrum. Due to the small isolated $WSe_2$ area used in the DT study, the laser beam

could not completely avoid the heterobilayer region, which results in the artifact of small positive signal at MoSe$_2$ exciton energy on the WSe$_2$ sample region.

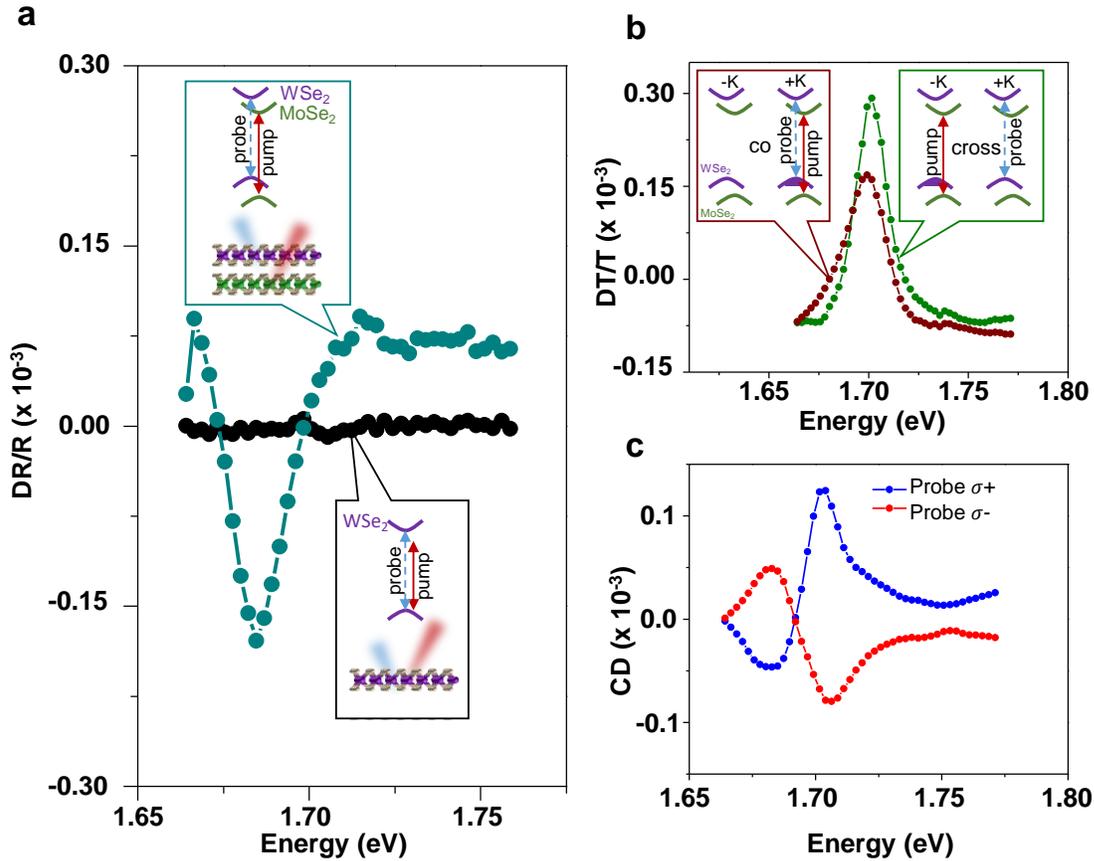

**Figure 2| Interlayer hole spin-valley polarization transfer. a,** Non-degenerate differential reflection (DR) of a MoSe$_2$-WSe$_2$ heterostructure, and an isolated WSe$_2$ region on the sample. When pumping on the lower energy MoSe$_2$ exciton resonance (1.621 eV), there is a strong DR response corresponding to the WSe$_2$ exciton (dark cyan), whereas the isolated WSe$_2$ monolayer shows negligible DR response (black). Co-circularly polarized pump and probe is shown. The insets depict the pump-probe scheme. The pump is shown as a solid red line, and the probe is the dashed blue line. DR data were measured at 50 K. **b,** Co- (burgundy) and cross- (green) circularly polarized DT spectra of the WSe$_2$ exciton resonances, when pumping the low energy MoSe$_2$ exciton resonance at 1.621 eV. The insets show the pump and probe scheme, where the band filling of the WSe$_2$ valence is shown. The line shapes are discussed in the text. **c,** Pump-induced circular dichroism (CD) of the WSe$_2$ exciton resonances when pumping MoSe$_2$ at 1.621 eV. CD highlights the differences between co- and cross-polarized DT responses. As expected, the sign of the CD response flips with probe (or pump) helicity.

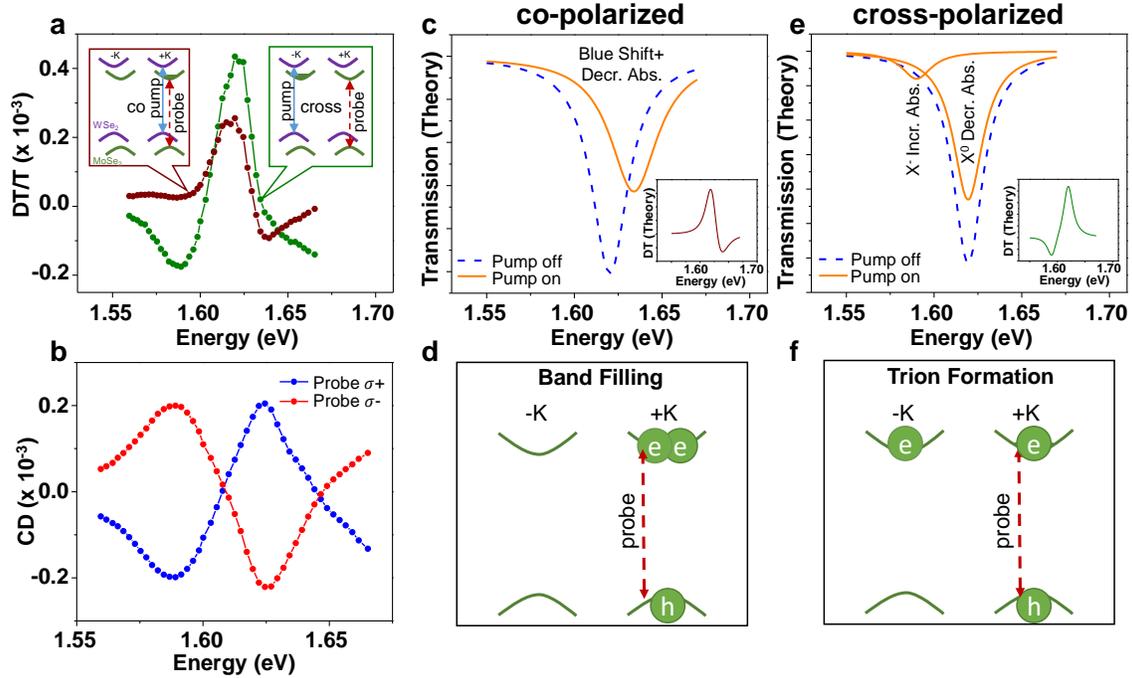

**Figure 3| Interlayer electron spin-valley polarization transfer. a,** Co- (burgundy) and cross- (green) circularly polarized DT spectra of the MoSe$_2$ exciton resonances, when pumping the higher energy WSe$_2$ exciton resonance at 1.710 eV. The insets show the pump and probe scheme, where the pump is the solid blue line, and the probe is the dashed red line. Following the interlayer transfer of photo-excited electrons from WSe$_2$ to MoSe$_2$, spin-valley polarized electrons are pumped into the MoSe$_2$ layer. The band filling of the MoSe$_2$ conduction band is shown. **b,** The pump-induced circular dichroism (CD) of the MoSe$_2$ exciton resonances when pumping WSe$_2$ at 1.710 eV flips sign with probe (or pump) helicity. **c-f** Theoretical explanations of the DT line shapes. **c-d,** For co-polarized pump and probe, the polarized electrons populate the same valley that the probe measures. The dominant effect is a band filling effect, so that when the pump is on (orange curve in **c**), the resonance is blue shifted and the exciton absorption is partially saturated, relative to the pump off case (blue dashed curve in **c**). In this co-polarized configuration, a charged exciton cannot form due to Pauli blocking. **e-f,** For cross-polarized pump and probe, the polarized electrons populate the opposite valley that the probe measures. The dominant effect is charged exciton (X$^-$) formation, so that when the pump is on (orange curve in **e**), the transmission is decreased at the X$^-$ resonance, and increased at the neutral exciton (X$^0$) resonance, relative to the pump off case (blue dashed curve in **e**). The insets of **c** and **e** show the difference between the modelled pump on (orange) and pump off (blue dashed) curves, corresponding to the theoretical DT spectra.

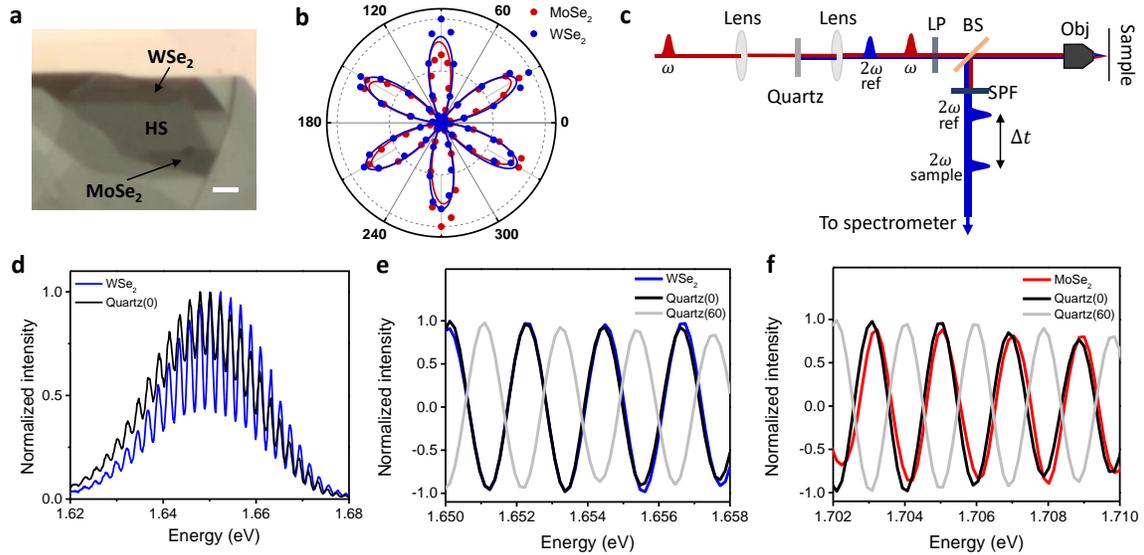

**Supplementary Figure 1| Second-harmonic generation measurements. a,** MoSe$_2$-WSe$_2$ heterostructure on sapphire from main text. The scale bar is 2 µm. **b,** SHG intensity parallel to the excitation polarization as a function of crystal angle for the monolayer MoSe$_2$ (red) and WSe$_2$ (blue) regions shown in **a**. The peaks of the lobes correspond to the armchair axes of the crystal. The relative angle between the lobe maxima is 1 ± 1°, meaning that the twist angle is close to 0° or 60°. **c,** Schematic of the phase-resolved SHG setup. LP, linear polarizer; BS, 50/50 beam splitter; SPF, short-pass filter; Obj, 50X objective. **d,** Second-harmonic interference spectrum from monolayer WSe$_2$ (blue) and from the front surface of z-cut quartz at 0° orientation (black). **e,** Close-up of the extracted SHG interference fringes for WSe$_2$ (blue) and two orientations of the quartz reference (black and gray), showing the SHG phase matches well with quartz(0) orientation compared to quartz(60). **f,** Same as in **e** but with MoSe$_2$ (red). This comparison shows the MoSe$_2$ phase agrees well with quartz(0) phase. Thus, the twist angle is near 0°.

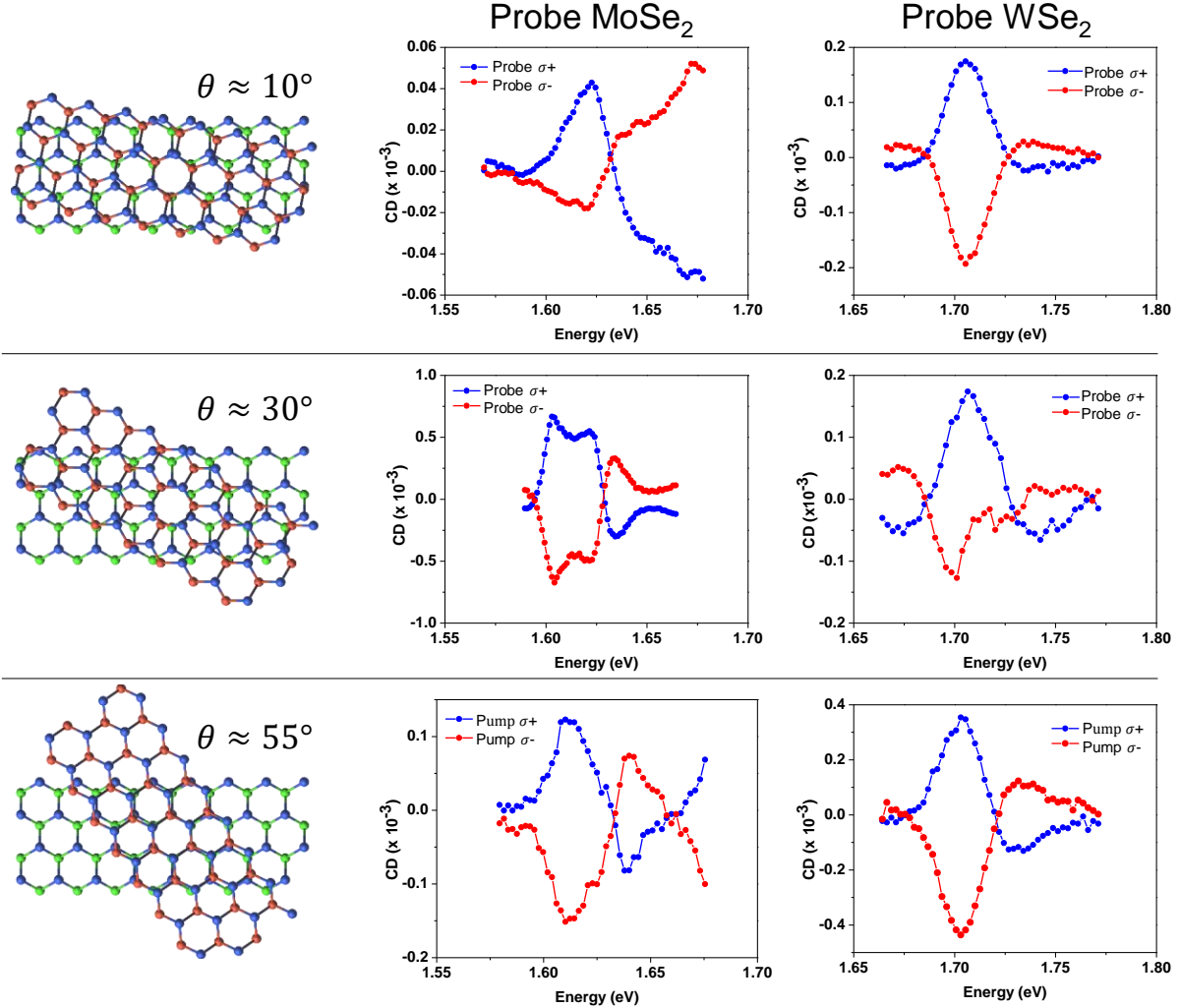

**Supplementary Figure 2| Additional CD measurements on SiO$_2$ substrate samples.** Each row corresponds to a different heterostructure sample with a different twist angle. The first column is a cartoon depicting the twist angle for each heterostructure. The blue atoms represent the selenium atoms and the red and green atoms represent the tungsten and molybdenum atoms, respectively. The second column shows the CD response from the MoSe$_2$ when pumping the WSe$_2$ near 1.71 eV. The third column shows the CD response from the WSe$_2$ when pumping the MoSe$_2$ near 1.63 eV. The pump and probe powers were in the 20-60 µW range. The labels for the data correspond to the pump and probe helicities respectively. For $\theta = 55°$, CD is shown varying the pump helicity instead of the probe helicity.

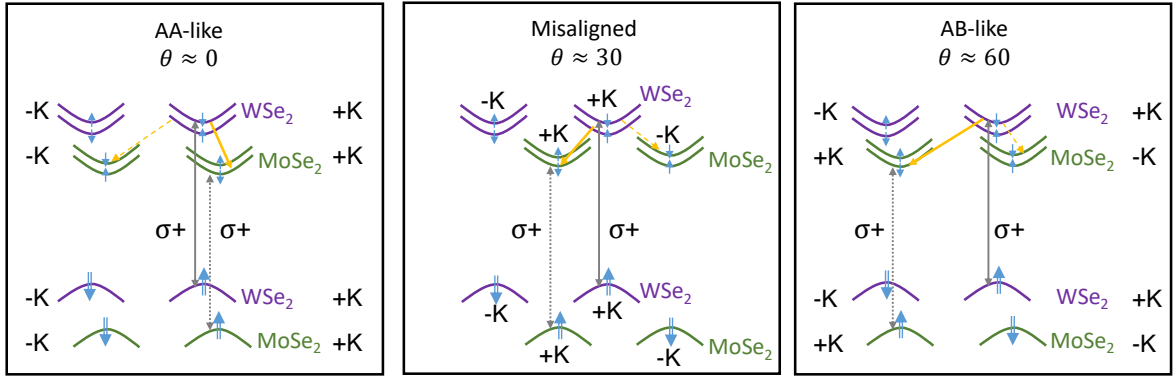

**Supplementary Figure 3| Spin-valley transfer in twisted heterostructures.** Electronic band structure for $MoSe_2$-$WSe_2$ heterostructures with different twist angles ($\theta$) showing the spin split conduction bands. The $WSe_2$ bands are shown in purple and the $MoSe_2$ bands are green. The blue arrows depict the real spin for electrons in the conduction bands and holes in the valence band. The solid gray line depicts a pump laser exciting the +K valley of $WSe_2$. The dotted gray line depicts a probe laser probing the +K valley of $MoSe_2$. The yellow arrows show the interlayer transfer processes that conserve real electron spin. The solid yellow arrow shows transfer to the lowest energy $MoSe_2$ conduction band, whereas the dashed yellow arrow shows the transfer to the higher energy $MoSe_2$ conduction band. The CD data for different twist angles (Supplementary Figure 5) indicates that the spin transfer process is dominated by transfer to the lowest energy band for all twist angles (the transfer process shown by the solid yellow arrow). The effects arising from the upper conduction band are discussed below.

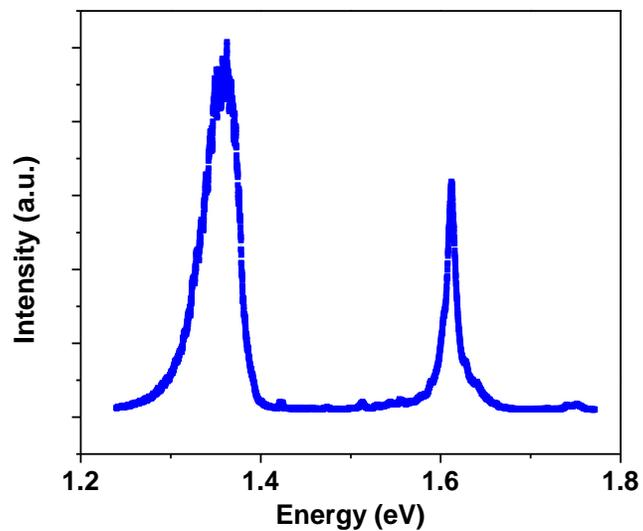

**Supplementary Figure 4| Heterostructure photoluminescence.** Photoluminescence spectrum of the heterostructure from the main text, recorded at 30 K with 30 µW excitation at 660 nm. Interlayer and MoSe$_2$ exciton peaks are observed centered at 1.355 eV and 1.612 eV respectively. The intralayer WSe$_2$ PL is negligible due to quenching.

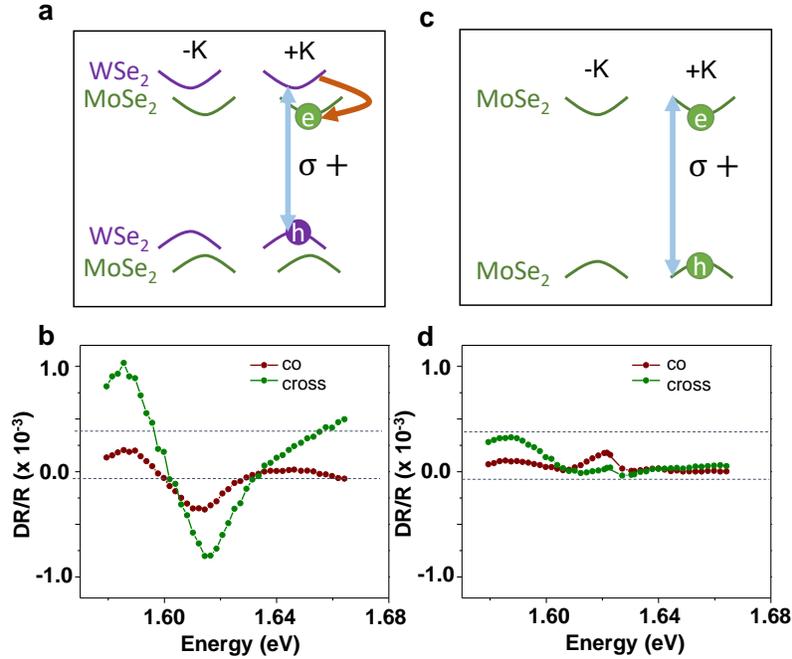

**Supplementary Figure 5| Heterostructure enhanced spin-valley polarization in MoSe$_2$. a,** In a heterostructure, +K spin-valley polarized electrons and hole are optically injected in the WSe$_2$ layer with circularly polarized light. The electron transfers to the +K conduction valley of the MoSe$_2$. **b,** DR spectra of the MoSe$_2$ exciton resonances when pumping WSe$_2$ at 1.687 eV, for co- (burgundy) and cross- (green) polarized pump and probe. **c,** In a monolayer MoSe$_2$ region, +K spin-valley polarized electrons and holes are directly optically injected in the MoSe$_2$ layer with circularly polarized light. **d,** DR spectra of the MoSe$_2$ exciton resonances when pumping at the WSe$_2$ resonance (1.687 eV) on an isolated MoSe$_2$ monolayer region for co-(burgundy) and cross- (green) polarized pump and probe. A comparison of **b** and **d** shows that the DR signal and the pumped-induced CD in **b** are primarily due to the interlayer electron transfer from the resonantly excited WSe$_2$ layer. Dashed lines are guides to the eye. DR measurements were performed at 50 K.

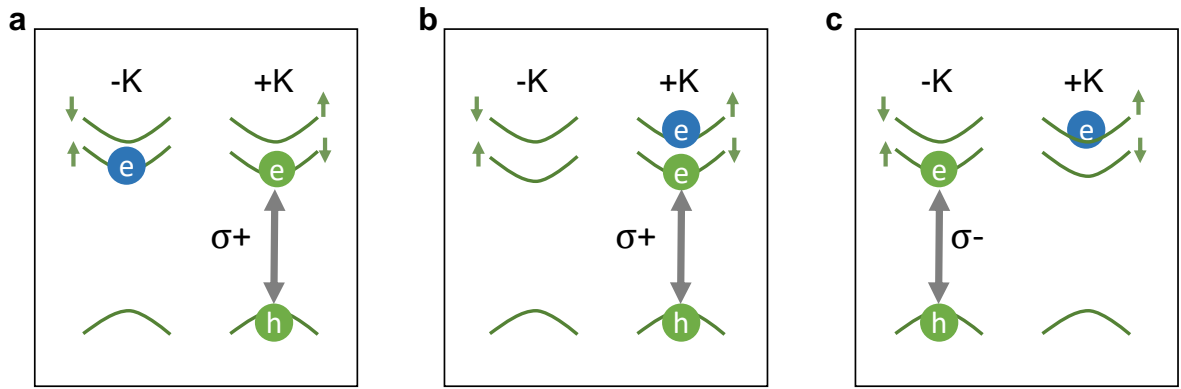

**Supplementary Figure 6| MoSe$_2$ charged exciton configurations. a,** The lowest energy negatively charged exciton has an intervalley configuration and consists of one electron in the lowest conduction band of one valley (-K shown) and the exciton in the opposite valley (+K shown). **b-c,** The higher energy charged exciton has the extra electron in the upper conduction band, which can have either an intravalley configuration (**b**) or an intervalley configuration (**c**). Note that the resonance energies for absorption or emission for all configurations are approximately equal, as discussed in S3. The arrow depicts the real spin of the electrons in each band.

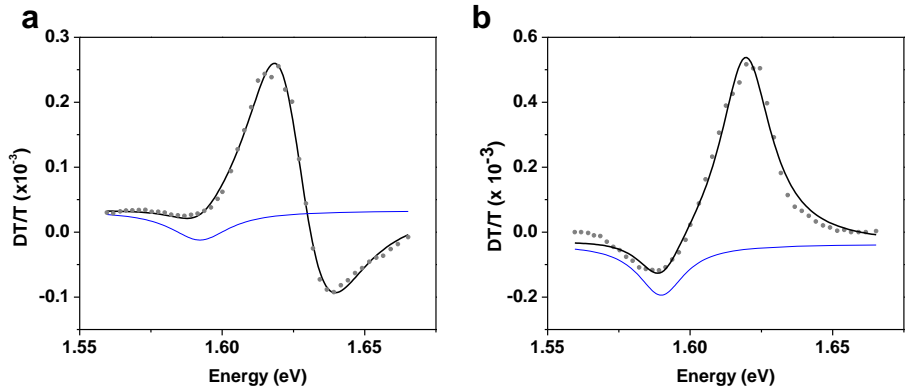

**Supplementary Figure 7| Estimating the spin-valley polarization in MoSe$_2$.** Detailed fitting to the data shown in Fig. 3a from the main text when pumping the WSe$_2$ heterostructure resonance near 1.710 eV, and probing the MoSe$_2$. We note that although we are plotting the same data as shown in Fig. 3a, a different aspect ratio is used to highlight the dip feature. By fitting the co- (**a**) and cross- (**b**) polarized DT spectra to a sum of three and two independent Lorentzians respectively, we compare the relative signal from the negative (dip) feature (shown in blue) centered near 1.59 eV, which corresponds to the energy of the negatively charged MoSe$_2$ exciton (X$^-$). Curve fitting reveals that the ratio of the areas for this dip feature is approximately 1: 0.37, yielding a spin-valley polarization of $(1-.37)/(1+.37) = 46\%$.

**Supplementary Note 1:**

**Measuring the Monolayer Crystal Axes by Polarization-Resolved and Phase-Sensitive Second-Harmonic Generation**

To determine the relative crystal orientation between the monolayers in the heterostructure, we performed polarization-resolved and phase-sensitive second-harmonic generation (SHG) measurements. Here, we detail the measurement procedure for the heterostructure in the main text (Supplementary Figure 1a).

First, the orientation of the armchair axes of each monolayer was determined using co-linear polarization-resolved SHG as shown in Supplementary Figure 1b[1-4]. Since this technique is not sensitive to the phase of the SHG, it only determines the twist angle up to 60°. In this case, the relative angle between the lobe maxima is 1 ± 1°, which means the twist angle is close to 0° or 60°.

To resolve this ambiguity, we performed a SHG spectral interference experiment modelled after Ref 5-7. An optical parametric amplifier, pumped by an amplified mode-locked Ti:sapphire laser, launched tunable ~200 fs pulses at $\omega$ (typically ~0.83 eV) into the setup shown in Supplementary Figure 1c. The excitation was focused onto a z-cut quartz reference, which was oriented to generate horizontally polarized $2\omega$ reference pulses (collinear to and co-polarized with the $\omega$ pulses). A time delay (Δt) was generated as the $\omega$ and $2\omega$ pulses travelled through the dispersive optics in the setup. A 50X objective focused the pulses onto the sample, generating an additional $2\omega$ sample signal (time-delayed from the $2\omega$ reference). The sample was oriented so an armchair axis, determined by the 6-fold pattern, was parallel to the $2\omega$ reference and $\omega$ polarization. The back-reflected signals were diverted with a 50/50 beamsplitter through a 1 $\mu$m short-pass filter, into a spectrometer, and detected by a Si CCD. The time-delayed $2\omega$ sample and reference signals produce spectral interference as shown in Supplementary Figure 1d. The period of the fringes is set by the time delay between the pulses (~2 ps) and the phase of the interference fringes is determined in part by the phase of SHG generated by the sample[5]. Therefore, this technique can distinguish between armchair axes 60° apart, which shows up as a $\pi$ phase shift in the interference spectrum. The fringes can be separated from the broad background SHG through Fourier transforms, as shown in Supplementary Figure 1e-f[6,7].

Due the different resonances of $MoSe_2$ and $WSe_2$, it can be difficult to find a single excitation energy for direct comparison of the relative phase. To circumvent this issue, we also performed the interference experiment using a reference sample of z-cut quartz. When focusing on the front surface of the quartz, the allowed second-order susceptibility elements are the same as those of the monolayer $MX_2$. Thus, as before, we oriented the quartz with its "armchair" axis parallel to the excitation and reference pulses to get a similar interference spectrum (Supplementary Figure 1d, black). This was done for two orientations of the quartz, 60° apart. The $WSe_2$ and $MoSe_2$ interference experiments were then compared with the quartz experiment (Supplementary Figure 1e-f, respectively). In the case of the sample shown in Supplementary Figure 1, both the $WSe_2$ and $MoSe_2$ SHG fringes align well with the quartz(0) direction, which confirms the orientation is close to AA-like stacking.

**Supplementary Note 2:**

**Effect of Twist Angle on the CD Response**

Additional pump-induced circular dichroism (CD) measurements were repeated in the reflection geometry on three $MoSe_2$-$WSe_2$ heterostructures on $SiO_2$ substrates. The twist angles of these samples were not aligned ($\theta \neq 0$); however, we observed that both the sign and qualitative line shape (sign and amplitude) of the CD response were not significantly affected by the twist angle (Supplementary Figure 2).

It has been previously established that the heterostructure twist angle affects how the $MoSe_2$ and $WSe_2$ valleys line up in momentum space[4,8] (Supplementary Figure 3). Therefore, the independence of the CD response with respect to twist angle suggests that the interlayer spin transfer process does not require the valleys to line up in momentum space. Thus, we conclude that the interlayer CD response we measure arises from real spin conservation during the interlayer transfer process (as opposed to valley pseudospin conservation). In the simplest model, spin-polarized carriers transfer between layers and then relax into the lowest energy state in each layer while maintaining their spin polarization as depicted in Supplementary Figure 3. We note that this conservation of real spin polarization also leads to a valley polarization in each layer due to the spin-valley locking effect[9].

**Supplementary Note 3:**

**Degenerate DT Line Shape and Doping Effects**

In this note, we examine the degenerate DT spectrum (Fig. 1c) and discuss the effects of unintentional doping. In the heterostructure region, the MoSe$_2$ (WSe$_2$) resonance is fit by a difference of two Lorentzians which reveals a weaker dip feature centered at 1.594 eV (1.675 eV) and stronger peak feature centered at 1.625 eV (1.705 eV). For each resonance, the ~20-30 meV difference between the peak and dip is consistent the reported binding energies for charged excitons. We therefore attribute the positive DT signal peaks to reduced neutral exciton absorption and the negative DT signal dips to increased charged exciton absorption. Both effects can be explained by pump induced photo-doping which increases the oscillator strength for charged excitons and reduces the oscillator strength for neutral excitons.

We note that the low energy pump induced absorption feature is stronger for the MoSe$_2$ layer. We attribute this difference between the WSe$_2$ and MoSe$_2$ response to the different species of charged exciton for each layer. As discussed in the main text, the WSe$_2$ pump induced absorption arises from formation of positively charged excitons ($X^+$), whereas the MoSe$_2$ pump induced absorption arises from negatively charged excitons ($X^-$). One possibility is that the oscillator strength for the $X^+$ in WSe$_2$ is weaker than that of the $X^-$ in MoSe$_2$, which is consistent with previous studies showing that the $X^+$ in WSe$_2$ has weaker PL[10]. We also note that PL measurements indicate that our MoSe$_2$ is weakly electron doped, whereas the WSe$_2$ is nearly intrinsic. However, since we are performing a differential measurement, we expect the effect of the background unpolarized doping to be small.

**Supplementary Note 4:**

**Effect of the Upper Conduction Band in MoSe$_2$**

In the main text, we use the relative amplitudes of the DT responses near the negatively charged exciton ($X^-$), to estimate the electron spin polarization in the MoSe$_2$ layer. The lowest energy $X^-$ corresponds to an intervalley configuration where, for example, one electron is in the lowest conduction band of one valley (-K), and the exciton is in the other valley (+K), as depicted in Supplementary Figure 6a. An intervalley trion with two electron in the same valley is not allowed due to Pauli blocking. Therefore, a -K valley electron in the lower conduction band can only induce $\sigma+$ polarized optical absorption at the $X^-$ resonance. The conduction band splitting has been calculated to be on order of 20 meV[11].

However, there is also the possibility of an $X^-$ composed of an electron in the upper conduction band which can either be in an intravalley (Supplementary Figure 6b) or intervalley (Supplementary Figure 6c) configuration and couples to a $\sigma+$ or $\sigma-$ polarized photon. The photon energy is given by the $X^-$ energy minus the upper conduction band electron energy, close to the lowest energy $X^-$ optical absorption. We note that: (1) The upper conduction band electrons have higher energies, so they will relax to the lower conduction band with the same spin[12]. Thus, in the steady state, the upper conduction band population is expected to be small. (2) A small electron population in the upper conduction band does not significantly affect our results. For example, if there is a small population of electrons in the upper conduction band, it increases the oscillator strength both for intravalley and intervalley charged excitons as shown in Supplementary Figure 6b-c. This population would lead to increased absorption for both co-polarized ($DT/T(Co)$) and cross-polarized ($DT/T(Cross)$) signals, which contributes to $DT/T(Cross) + DT/T(Co)$. But to lowest order, these two contributions are equal in magnitude, so they give rise to a negligible CD response which is proportional to the difference, $DT/T(Cross) - DT/T(Co)$. In this case, the value $\rho = \frac{DT/T(Cross) - DT/T(Co)}{DT/T(Cross) + DT/T(Co)}$ corresponds to a lower bound for the spin-valley polarization of the lower conduction band electron in the MoSe$_2$ layer.

**Supplementary References:**